\documentclass[a4paper,10pt,twocolumn,oneside,conference]{IEEEtran}
\IEEEoverridecommandlockouts
\usepackage{cite}
\usepackage{amsmath,amssymb,amsfonts}
\usepackage{graphicx}
\usepackage{textcomp}
\usepackage{xcolor}
\usepackage{makecell}
\usepackage{multicol}
\usepackage{siunitx}
\usepackage{booktabs}
\usepackage{xfrac}
\usepackage{algorithm}
\usepackage{algpseudocode}
\usepackage[nolist]{acronym}
\usepackage{tikz}
\usepackage{pgfplots}
\pgfplotsset{width=7cm,compat=1.3}
\usetikzlibrary{shapes,arrows,positioning,calc, matrix,spy,patterns}
\usetikzlibrary{decorations.pathreplacing,calligraphy}
\usetikzlibrary{colorbrewer}
\usetikzlibrary{fit}
\usetikzlibrary{external}
\usepgfplotslibrary{colorbrewer}

\definecolor{cb-1}{HTML}{4477AA}
\definecolor{cb-2}{HTML}{EE6677}
\definecolor{cb-3}{HTML}{228833}
\definecolor{cb-4}{HTML}{CCBB44}
\definecolor{cb-5}{HTML}{66CCEE}
\definecolor{cb-6}{HTML}{AA3377}
\definecolor{cb-7}{HTML}{BBBBBB}

\pgfplotscreateplotcyclelist{cb list}{
	cb-1,cb-2,cb-3,cb-4,cb-5,cb-6,cb-7
}

\definecolor{kit-green100}{rgb}{0,.59,.51}
\definecolor{kit-green70}{rgb}{.3,.71,.65}
\definecolor{kit-green50}{rgb}{.50,.79,.75}
\definecolor{kit-green30}{rgb}{.69,.87,.85}
\definecolor{kit-green15}{rgb}{.85,.93,.93}
\definecolor{KITgreen}{rgb}{0,.59,.51}

\definecolor{KITpalegreen}{RGB}{130,190,60}
\colorlet{kit-maigreen100}{KITpalegreen}
\colorlet{kit-maigreen70}{KITpalegreen!70}
\colorlet{kit-maigreen50}{KITpalegreen!50}
\colorlet{kit-maigreen30}{KITpalegreen!30}
\colorlet{kit-maigreen15}{KITpalegreen!15}

\definecolor{KITblue}{rgb}{.27,.39,.66}
\definecolor{kit-blue100}{rgb}{.27,.39,.67}
\definecolor{kit-blue70}{rgb}{.49,.57,.76}
\definecolor{kit-blue50}{rgb}{.64,.69,.83}
\definecolor{kit-blue30}{rgb}{.78,.82,.9}
\definecolor{kit-blue15}{rgb}{.89,.91,.95}

\definecolor{KITyellow}{rgb}{.98,.89,0}
\definecolor{kit-yellow100}{cmyk}{0,.05,1,0}
\definecolor{kit-yellow70}{cmyk}{0,.035,.7,0}
\definecolor{kit-yellow50}{cmyk}{0,.025,.5,0}
\definecolor{kit-yellow30}{cmyk}{0,.015,.3,0}
\definecolor{kit-yellow15}{cmyk}{0,.0075,.15,0}

\definecolor{KITorange}{rgb}{.87,.60,.10}
\definecolor{kit-orange100}{cmyk}{0,.45,1,0}
\definecolor{kit-orange70}{cmyk}{0,.315,.7,0}
\definecolor{kit-orange50}{cmyk}{0,.225,.5,0}
\definecolor{kit-orange30}{cmyk}{0,.135,.3,0}
\definecolor{kit-orange15}{cmyk}{0,.0675,.15,0}

\definecolor{KITred}{rgb}{.63,.13,.13}
\definecolor{kit-red100}{cmyk}{.25,1,1,0}
\definecolor{kit-red70}{cmyk}{.175,.7,.7,0}
\definecolor{kit-red50}{cmyk}{.125,.5,.5,0}
\definecolor{kit-red30}{cmyk}{.075,.3,.3,0}
\definecolor{kit-red15}{cmyk}{.0375,.15,.15,0}

\definecolor{KITpurple}{RGB}{160,0,120}
\colorlet{kit-purple100}{KITpurple}
\colorlet{kit-purple70}{KITpurple!70}
\colorlet{kit-purple50}{KITpurple!50}
\colorlet{kit-purple30}{KITpurple!30}
\colorlet{kit-purple15}{KITpurple!15}

\definecolor{KITcyanblue}{RGB}{80,170,230}
\colorlet{kit-cyanblue100}{KITcyanblue}
\colorlet{kit-cyanblue70}{KITcyanblue!70}
\colorlet{kit-cyanblue50}{KITcyanblue!50}
\colorlet{kit-cyanblue30}{KITcyanblue!30}
\colorlet{kit-cyanblue15}{KITcyanblue!15}

\definecolor{cb-1}{HTML}{4477AA}
\definecolor{cb-2}{HTML}{EE6677}
\definecolor{cb-3}{HTML}{228833}
\definecolor{cb-4}{HTML}{CCBB44}
\definecolor{cb-5}{HTML}{66CCEE}
\definecolor{cb-6}{HTML}{AA3377}
\definecolor{cb-7}{HTML}{BBBBBB}

\pgfplotscreateplotcyclelist{cb list}{
	cb-1,cb-2,cb-3,cb-4,cb-5,cb-6,cb-7
}

\let\j\relax
\newcommand{\j}{\mathrm{j}}

\newcommand*{\Tcp}{T_{\mathrm{cp}}}
\newcommand*{\Ncp}{N_{\mathrm{cp}}}
\newcommand*{\Gp}{G_{\mathrm{P}}}
\newcommand*{\rect}{\mathrm{rect}}
\newcommand*{\e}{\mathrm{e}}
\newcommand*{\fDh}{f_{\mathrm{D},h}}
\newcommand*{\tah}{\tilde{a}_h}

\newcommand*{\RDM}{\mathrm{RDM}}
\newcommand*{\Ts}{T_{\mathrm{s}}}

\usepackage{xcolor}
\usepackage{amsmath}
\usepackage{amssymb}
\usetikzlibrary{calc}
\usetikzlibrary{positioning}
\usetikzlibrary{patterns}
\usetikzlibrary{decorations.pathreplacing,calligraphy}

\tikzset{
	lsblock/.style={rectangle, thick, draw, minimum width=1.2cm, minimum height=0.6cm, rounded corners=1.6mm, font=\small, align=center}
}
\tikzstyle{surround} = [rectangle, rounded corners, draw=KITred, inner sep=0.1cm, dashed, thick]

\def\BibTeX{{\rm B\kern-.05em{\sc i\kern-.025em b}\kern-.08em
    T\kern-.1667em\lower.7ex\hbox{E}\kern-.125emX}}

\begin{document}
\begin{acronym}[TROLOLO]
    \acro{6G}{sixth generation}
    \acro{CC}{coherent compensation}
    \acro{CP}{cyclic prefix}
    \acro{FD}{frequency domain}
    \acro{FFT}{fast Fourier transform}
    \acro{ICI}{inter-carrier interference}
    \acro{IFFT}{inverse fast Fourier transform}
    \acro{ISAC}{integrated sensing and communications}
    \acro{ISI}{inter-symbol interference}
    \acro{MTCC}{multi target coherent compensation}
    \acro{OFDM}{orthogonal frequency division multiplexing}
    \acro{RCS}{radar cross section}
    \acro{RDM}{range-Doppler matrix}
    \acro{SINR}{signal-to-interference-and-noise ratio}
    \acro{TD}{time domain}
    \acro{TOI}{target of interest}
\end{acronym}

\title{
Integrated Long-range Sensing and Communications in Multi Target Scenarios using CP-OFDM
\thanks{This work has received funding from the German Federal Ministry of Education and Research (BMBF) within the projects Open6GHub (grant agreement 16KISK010) and KOMSENS-6G (grant agreement 16KISK123 and gran agreement  16KISK112K).}
}

\author{\IEEEauthorblockN{Benedikt$\,$Geiger*,$\,$Silvio Mandelli\textsuperscript{\dag},$\,$Marcus$\,$Henninger\textsuperscript{\dag},$\,$Daniel$\,$Gil$\,$Gaviria*,$\,$Charlotte$\,$Muth*,$\,$Laurent$\,$Schmalen*}
\IEEEauthorblockA{*Communications Engineering Lab (CEL), Karlsruhe Institute of Technology (KIT), 76187 Karlsruhe, Germany\\
\textsuperscript{\dag}Nokia Bell Labs Stuttgart, 70469 Stuttgart, Germany\\
Email: \texttt{benedikt.geiger@kit.edu}}
}

\maketitle

\begin{abstract}
6G communication systems promise to deliver sensing capabilities by utilizing the \ac{OFDM} communication signal for sensing. However, the cyclic prefix inherent in \ac{OFDM} systems limits the sensing range, necessitating compensation techniques to detect small, distant targets like drones. In this paper, we show that state-of-the-art coherent compensation methods fail in scenarios involving multiple targets, resulting in an increased noise floor in the radar image. Our contributions include a novel multi target coherent compensation algorithm and a generalized signal-to-interference-and-noise ratio for multiple targets to evaluate the performance. Our algorithm achieves the same detection performance at long distances requiring only 3.6\% of the radio resources compared to classical \ac{OFDM} radar processing. This enables resource efficient sensing at long distances in multi target scenarios with legacy communications-only networks.
\end{abstract}

\acresetall
\section{Introduction}
By leveraging the communication signal also for sensing, \ac{ISAC} has the potential to transform the \ac{6G} of mobile communication networks, enabling the detection of passive objects in a radar-like manner~\cite{wild_6g_2023}. \ac{ISAC} promises reduced hardware complexity, higher spectral efficiency as well as energy efficiency compared to two separate systems, and novel use cases such as drone detection, which is a rapidly growing market~\cite{shatov_joint_2024, TR_22_837}.

\begin{figure}
    \centering
    \footnotesize
    \input{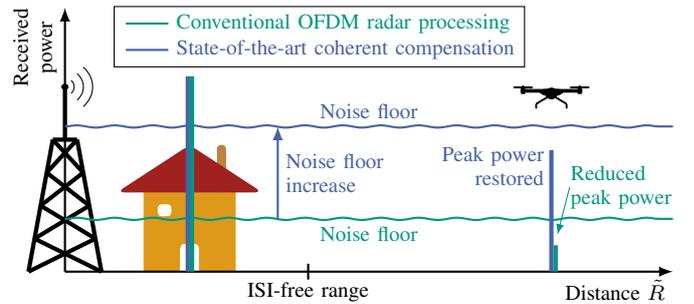}
    \vspace{-0.6cm}
    \caption{Scenario under investigation: Long-range drone detection in a multi target scenario using a communication-centric \acs{CP}-\acs{OFDM}-\acs{ISAC} system. The drone remains undetected using both algorithms. This highlights the need for a novel long-range sensing algorithm that works robustly in multi target scenarios. Conventional \ac{OFDM} radar processing suffers from a reduced peak power. The state-of-the-art \ac{CC}, which increases the signal power of distant targets, disproportionately increases the interference-and-noise floor, further preventing drone detection.
    }
    \label{fig:ch1:scenario}
    \vspace{-0.3cm}
\end{figure}

Like the current versions of cellular standards, \ac{6G}
\ac{ISAC} systems are expected to utilize \ac{OFDM} because it allows for efficient equalization and channel estimation in the \ac{FD} using the \ac{FFT}. To satisfy the periodicity requirement of the \ac{FFT} at the communication receiver, a \ac{CP}, which is longer than the delay spread of the communication channel, is added to the transmit symbols. In theory, the delay spread is infinite, but in practice, the belated reflections tend to be very weak, and the delay spread of the communications channel is limited to the relevant, i.e., strong, taps. While weak reflections exceeding the duration of the \ac{CP} do not significantly degrade communication, \ac{ISAC} systems should be capable of detecting objects that exceed the \ac{CP} duration. However, if the delay of an object exceeds the \ac{CP} duration, part of the reflection spills into the consecutive \ac{OFDM} symbol leading to \ac{ISI} and \ac{ICI}. As a result, the range corresponding to the \ac{CP} duration is often referred to as \ac{ISI}-free range~\cite{nuss_limitations_2018}, which, for example, is only $\SI{343}{\metre}$ in case of a normal \ac{CP} duration at $\SI{30}{kHz}$ sub-carrier spacing~\cite{TR_38_211}. More critically, the portion of the reflection that overlaps with the consecutive \ac{OFDM} symbol is not present in the actual \ac{OFDM} symbol, reducing the target's peak power in the \ac{RDM}, which could result in a missed detection, as visualized in the distant drone detection scenario in Fig.~\ref{fig:ch1:scenario}.

To address this limitation, time domain channel estimation can be used to detect targets beyond the \ac{ISI}-free range~\cite{oliari_ofdm_2024}. However, this method significantly increases computational complexity compared to conventional \ac{FFT}-based processing~\cite{oliari_ofdm_2024}. Alternatively, Wang et al.\cite{wang_coherent_2023} proposed a \ac{CC} algorithm that increases the peak power of the distant target by adding the current \ac{OFDM} symbol to the previous one. While this method operates in the frequency domain with only a slight increase in complexity, it introduces significant \ac{ISI} in multi target scenarios, which can be much stronger than the resulting signal power increase, as shown in Fig.~\ref{fig:ch1:scenario}, making it impractical for real-world applications. Thus, \ac{ISAC} systems lack a low-complexity, long-range sensing algorithm that can robustly handle multi target scenarios.

In this paper, we address this gap by investigating sensing beyond the \ac{ISI}-free range in multi target scenarios using a \ac{CP}-\ac{OFDM}-\ac{ISAC} system. Our contributions include:
\begin{itemize}
    \item We derive the image \ac{SINR} for scenarios where targets exceed the \ac{ISI}-free range.
    \item We determine the maximum detection range of distant targets with a small \ac{RCS} using a communication-centric \ac{ISAC} system demonstrating the need for a long-range sensing algorithm.
    \item We demonstrate the \ac{ISI} issue of the state-of-the-art \ac{CC} method in multi target scenarios.
    \item We propose a novel \ac{MTCC} algorithm for robust long-range sensing in multi target scenarios.
    \item We verify all our results through simulations.
\end{itemize}

\section{System model}
\label{sec:ch2:system_model}
\begin{figure}
    \centering
    \input{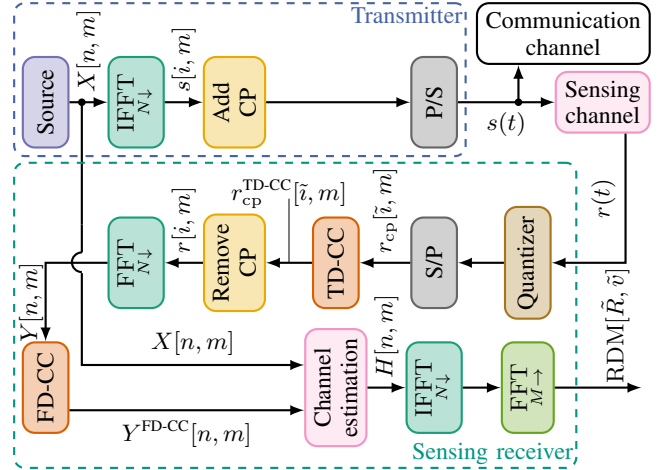}
    \vspace{-0.65cm}
    \caption{Block diagram of the \ac{OFDM} \ac{ISAC} signal processing architecture including the \ac{TD}- and \ac{FD}-\ac{CC} assuming an ideal radio frequency frontend.}
    \label{fig:ch2:System_setup}
    \vspace{-0.40cm}
\end{figure}

In this paper, we consider a monostatic \ac{OFDM}-\ac{ISAC} system in which a sensing receiver is co-located with the transmitter, e.g., a base station. 
The co-location of the transmitter and sensing receiver ensures synchronization and allows the sensing receiver to access the transmitted data.
The \ac{ISAC} system follows a communication-centric design and transmits information to mobile users while simultaneously sensing the environment based on the reflections of the transmit signal. Figure~\ref{fig:ch2:System_setup} shows a block diagram of the signal processing architecture that we detail in the following.

\subsection{Transmitter}
One \ac{OFDM} frame consists of $M$ \ac{OFDM} symbols, each using $N$ orthogonal sub-carriers with sub-carrier spacing $\Delta f = B/N = 1/(\Ts N) = 1/T$, where $B$, $\Ts$, and $T$ denote bandwidth, sample duration, and OFDM symbol duration, respectively. The transmitter modulates the constellation symbols $X[n,m]$ \mbox{($n = 1, \ldots, N, m = 1, \ldots, M$)} from a modulation alphabet~$\mathcal{A}$, e.g., QPSK, and adds a \ac{CP} to each symbol before transmitting the equivalent baseband signal~\cite{braun_ofdm_2014}
\begin{equation}
    s(t) \! = \! \frac{1}{\sqrt{N}} \! \sum_{m = 1}^{M} \sum_{n = 1}^{N} X[n,m] \mathrm{e}^{\mathrm{j} 2\pi n \Delta f (t \! - \! m T_0)} g(t \! \! - \! \! mT_0),
\end{equation}
where $T_0 \! = \! T \! + \! \Tcp$ is the \ac{OFDM} symbol duration incl. \ac{CP} and $g(t) \! = \! \mathrm{rect}_{[-\Tcp,T)}(t)$ is a rectangular function \mbox{$\rect_{[a,b)}(t)$}, which is $1$ if $ t \in [a,b)$ and $0$ otherwise.

\subsection{Sensing Channel Model}
Assuming a Swerling-0 model~\cite{richards_fundamentals_2014} and $H$ point targets each having a distance $R_h$, radial velocity $v_h$ and \ac{RCS} $\sigma_{\mathrm{RCS},h}$, the received sensing signal is~\cite{braun_ofdm_2014}
\begin{equation}
    r(t) = \sum_{h=1}^{H} a_h s(t - \tau_h) \mathrm{e}^{ \j 2 \pi \left( f_{\mathrm{D},h} t + \varphi_h \right) } + w(t),
\end{equation}
where amplitude $a_h$, delay $\tau_h$, and Doppler frequency $f_{\mathrm{D},h}$ of target $h$ are \cite{braun_ofdm_2014}
\begin{equation}
    a_h\! = \! \sqrt{ \frac{\sigma_{\mathrm{RCS},h} c_0^2 P_{\mathrm{Tx}} G_{\mathrm{Tx}} G_{\mathrm{Rx}}}{\left( 4 \pi \right)^3 R_h^4 f_c^2}}, \,
    \tau_h \! = \! 2 \frac{R_h}{c_0}, \,
    \fDh \! = \! 2 \frac{v_h f_{\mathrm{c}}}{c_0}.
\end{equation}
The parameters $P_{\mathrm{Tx}}$, $G_{\mathrm{Tx}}$, $G_{\mathrm{Rx}}$, $f_{\mathrm{c}}$, and $c_0$ represent transmit power, gain of the transmit and receive antenna, carrier frequency and speed of light, respectively. The phase~$\varphi_h$ is uniformly sampled from the interval $[0, 2 \pi)$ and the thermal noise is $w(t) \sim \mathcal{CN} \left( 0, \sigma_{\mathrm{Therm}}^2 \right)$, with noise power \mbox{$\sigma_{\mathrm{Therm}}^2 = F k B T_{\mathrm{Therm}}$}. Here, $k$ is Boltzmann's constant, $T_{\mathrm{Therm}}$ the equivalent noise temperature, and $F$ the noise figure of the receiver.

\subsection{Receiver}
\label{sec:ch2:receiver}
At the sensing receiver, the received signal $r(t)$ is filtered, sampled, quantized using $Q$ bits, and the \ac{CP} is removed. Next, the signal is transformed into the frequency domain by taking the \ac{FFT} along the $N$ sub-carriers. In the frequency domain, the channel is estimated by an element-wise division of the receive symbols by the transmit symbols%
\vspace{-0.1cm}
\begin{equation}
    H[n,m] = \frac{Y[n,m]}{X[n,m]} \odot H_{\mathrm{win}} [n,m],
    \label{eq:ch2:channel_estimation}
    \vspace{-0.1cm}
\end{equation}
where $\odot$ is the Hadamard product and $H_{\mathrm{win}}$ a windowing function to suppress sidelobes in the \ac{RDM}. Finally, the \ac{RDM} matrix is obtained by an orthonormal \ac{IFFT} along the sub-carriers, and an orthonormal \ac{FFT} along the OFDM symbols \cite{braun_ofdm_2014}
\vspace{-0.1cm}
\begin{equation}
    \mathrm{RDM}[\tilde{R},\tilde{v}] = \underset{M \rightarrow}{\mathrm{FFT}} \left\{ \underset{N \downarrow}{\mathrm{IFFT}} \left\{ H[n,m] \right\} \right\}.
    \label{eq:ch2:RDM}
    \vspace{-0.2cm}
\end{equation}

\begin{figure*}
    \centering
    \input{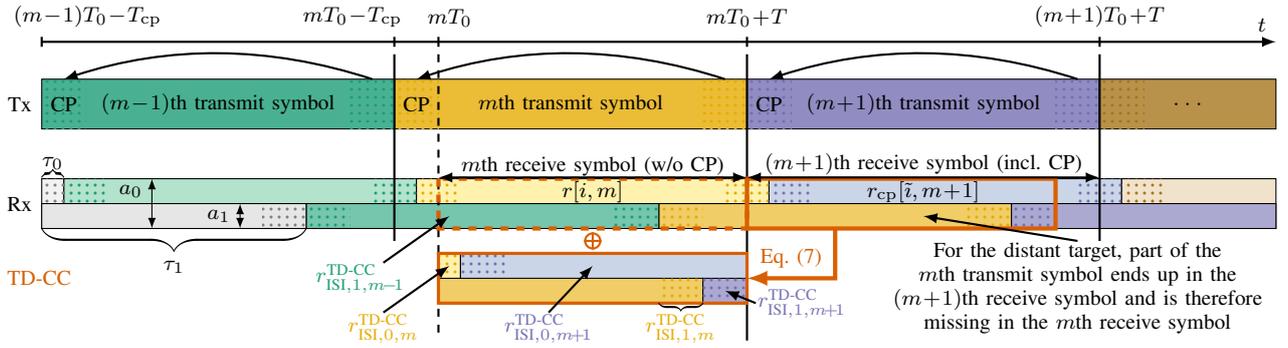}
    \vspace{-0.3cm}
    \caption{The first and the second row visualize the transmit and receive signal in the case of a close and distant reflection, i.e., target, having an amplitude and delay of $a_0,\! \tau_0$ and $a_1,\! \tau_1$, respectively. The third row and orange boxes visualize the proposed \ac{TD}-\ac{CC}: The $N$ samples following the $m$th receive symbol are added to the $m$th receive symbol. The resulting \ac{ISI} after \ac{TD}-\ac{CC} is labeled as in~(\ref{eq:ch2:r_ISI}). We assume a synchronized transmitter and receiver in our monostatic setup.~\cite{wild_6g_2023}}
    \label{fig:ch2:Add_compensation}
    \vspace{-0.4cm}
\end{figure*}

\subsection{Distant Targets}
\label{sec:ch2:target_exceeding_CP}
Since the \ac{CP} is optimized for communications and not for sensing, the delay of a \ac{TOI} can easily exceed the duration of the \ac{CP}. For a delay \mbox{$0 \leqslant \tau_h < T$}, the $h$th target's contribution to the $m$th received symbol after \ac{CP} removal is \cite{wang_coherent_2023}
\vspace{-0.4cm}
\begin{equation}
    \begin{gathered}
        \vspace*{-0.3cm}
        r_h[i,m] =  \frac{1}{\sqrt{N}} \left( \sum_{n=1}^{N} X[n,m] \e^{\j 2 \pi n \Delta f ( i \Ts \! - \! \tau_h)} g (i \Ts \! - \! \tau_h)  \right. \\
        \vspace{-1.3cm}
        + \! \left. \sum_{n=1}^{N} \! X[n,m \! - \! 1] \e^{\j 2 \pi n \Delta f ( i \Ts \! - \! \tau_h \! + \! T_0 ) } g (i \Ts \! - \! \tau_h \! + \! T_0) \right) \tah[m], %
    \end{gathered}
    \vspace{1.0cm}
\end{equation}
where \mbox{$i = 1, \ldots, N$} denotes the sample in the time domain and \mbox{$\tah[m] = a_h \exp\left( \j 2 \pi \fDh m T_0 \! + \! \j \varphi_h \right)$}~\cite{braun_ofdm_2014}. 

Figure~\ref{fig:ch2:Add_compensation} depicts the transmitted and received signals for two targets, one close and one distant, as well as the \ac{CC}. It can be observed in the second row of Fig.~\ref{fig:ch2:Add_compensation} that, if a delay exceeds $T_{\mathrm{cp}}$, part of the \mbox{$(m\!-\!1)$th} transmit symbol ends up in the \mbox{$m$th} receive symbol, leading to \ac{ISI}. Additionally, the \mbox{$m$th} receive symbol does not contain the complete \mbox{$m$th} transmit \ac{OFDM} symbol, leading to a reduced peak power and \ac{ICI}~\cite{wang_coherent_2023}. In~\cite{wang_coherent_2023}, analytical expressions for the peak power, \ac{ISI}, \ac{ICI}, and \ac{SINR} for the received symbols $Y$ in the \ac{FD} are derived. We expand these results and give in the following the \ac{SINR} in the \ac{RDM} because it is a crucial metric for detection. For the sake of conciseness, we present only the key ideas of the derivation. The element-wise symbol division to estimate the channel does not change the power if QPSK is used. For the sake of simplicity, we assume that QPSK symbols are transmitted as they outperform higher order modulation formats for sensing due to noise enhancement~\cite{braun_ofdm_2014}. The peak power experiences an integration gain $G_{\mathrm{P}} = MN$ due to the coherent addition in the (I)\ac{FFT}. In contrast, the \ac{ISI} and \ac{ICI} are uncorrelated and retain the same power in the \ac{RDM} as in the frequency domain due to Parseval's theorem. 
Our derived analytical expressions for the peak power, \ac{ISI}, and \ac{ICI} in the \ac{RDM} are summarized in Tab.~\ref{tab:ch2:power_level}.

\section{Multi Target Coherent Compensation}
In this section, we summarize and generalize the \ac{TD}-\ac{CC} proposed by \cite{wang_coherent_2023}, demonstrate its deficiency in a multi target scenario and propose our novel and robust \ac{MTCC}. 

    \begin{table*}
    \caption{Summary of the derived peak, \ac{ISI} and \ac{ICI} power of the $h$th target in the \ac{RDM} assuming a QPSK transmission where the derivation is based on~\cite{wang_coherent_2023}. 
    For brevity, we define $\Tcp^{\mathrm{min}} = \min (\Tcp, (\tau_h - \Tcp)^+ )$ and $(\cdot)^+ = \max(\cdot , 0)$.}
    \vspace{-0.35cm}
    \label{tab:ch2:power_level}
    \normalsize
    \begin{tabular}{ccccc}
        \toprule
        Name & \makecell{No CC}  & TD-CC & FD-CC & MTCC \\
        \midrule
        $ P_{\mathrm{Peak},h} $& $\Gp a_h^2 \left( 1 \! - \! \frac{(\tau_h - \Tcp)^+}{T} \right)^2$ & $\Gp a_h^2 \left( 1 \! + \! \frac{\min(\tau_h, \Tcp)}{T}\right)$ & $\Gp a_h^2$ & $\Gp a_h^2$ \\
        $ P_{\mathrm{ISI},h}$ & $a_h^2 \left( \frac{(\tau_h - \Tcp)^+}{T} \right)$ & $a_h^2$ & $a_h^2 \left(1 + \frac{\Tcp^{\mathrm{min}}}{T} \right)$ & $ a_h^2 \frac{(\tau_h \! - \! \Tcp)^+}{T}  \! + \! \min \! \left( \! a_h^2, \frac{\varepsilon}{\Gp} \! \right) \!\frac{T \! - \! (\tau_h \! - \! \Tcp)^+}{T} $  \\
        $ P_{\mathrm{ICI},h}$ & $a_h^2 \frac{ \left( T- \Tcp - \tau_h \right) \left( \tau_h - \Tcp \right)^+ }{T^2}$ & $0$ & $a_h^2 \frac{ \left( T - \Tcp^{\mathrm{min}} \right) \Tcp^{\mathrm{min}} }{T^2}$ & $2 a_h^2 \frac{ \left( T - \Tcp^{\mathrm{min}} \right) \Tcp^{\mathrm{min}} }{T^2}$ \\
        $P_{\mathrm{Therm}}$ & $\sigma^2_{\mathrm{Therm}}$ & $2 \sigma^2_{\mathrm{Therm}}$ & $2 \sigma^2_{\mathrm{Therm}}$ & $2 \sigma^2_{\mathrm{Therm}}$ \\
        \bottomrule
    \end{tabular}
    \vspace{-0.2cm}
    \end{table*}

\subsection{Coherent Compensation}
\label{sec:ch3:GCC}
The \ac{TD}-\ac{CC} algorithm proposed in \cite{wang_coherent_2023} is based on the observation that the \mbox{$m$th} receive \ac{OFDM} symbol must contain the complete \mbox{$m$th} transmit \ac{OFDM} symbol to have full peak power and to prevent \ac{ICI}. Figure~\ref{fig:ch2:Add_compensation} shows that part of the $m$th transmit symbol ends up in the \mbox{$(m\!+\!1)$th} receive symbol if $\tau_h > \Tcp$. Therefore, the \ac{TD}-\ac{CC} adds the samples of the \mbox{$(m\!+\!1)$th} received symbol belonging to the \mbox{$m$th} transmit symbol to the start of the \mbox{$m$th} receive symbol. However, this requires knowledge about the delay $\tau_{\mathrm{TOI}}$ of the \ac{TOI} which should be detected, to determine the number of samples to be added, resulting in a chicken-and-egg problem. Further, scenarios with multiple targets are not discussed in the original publication. Therefore, we generalize the \ac{TD}-\ac{CC} proposed in \cite{wang_coherent_2023} to ensure that the complete \mbox{$m$th} transmit symbol is in the compensated \mbox{$m$th} receive symbol restoring the full peak power and preventing \ac{ICI} for multiple targets with unknown delays $0 \leqslant \tau_h < T$. As depicted in Fig.~\ref{fig:ch2:Add_compensation}, we propose to add the $N$ samples following the \mbox{$m$th} receive symbol to the $m$th receive symbol, i.e., for \mbox{$\tilde{\imath}\!=\!1,\ldots, N$}
\begin{equation}
    r^{\text{TD-CC}}_{\mathrm{cp}}[\tilde{\imath}+\Ncp,m] = r_{\mathrm{cp}}[\tilde{\imath}+\Ncp,m] + r_{\mathrm{cp}}[\tilde{\imath},m+1].
    \label{eq:TD_CC}
\end{equation}
After \ac{TD}-\ac{CC}, the received reflection of the $h$th target without \ac{CP} becomes a cyclically shifted version of the transmit signal
\begin{equation}
    r^{\text{TD-CC}}_h[i,m] = \tah[m] \left( s[i \! \! + \! \! \frac{\tau_h}{T} \ \mathrm{mod} \ N,m] + r^{\text{TD-CC}}_{\mathrm{ISI},h}[i,m] \right) %
    \vspace{-0.15cm}
\end{equation}
and an \ac{ISI} term, which is
\begin{equation}
    \setlength{\jot}{-8pt}
    \begin{gathered}
        r^{\text{TD-CC}}_{\text{ISI},h}[i,m] = \underbrace{\sum_{n=1}^{N} X[n,m \! + \! 1] \e^{\j 2 \pi n \Delta f \left( i \Ts - \tau_h \right)} g (i \Ts \! - \! \tau_h)}_{\color{Dark2-C}r^{\text{TD-CC}}_{\text{ISI},h,m\!+\!1}}  \\
        + \! \! \underbrace{\sum_{n=1}^{N} \! \! X[n,m]  \e^{\j 2 \pi n \Delta f \left( i \Ts + \Tcp - \tau_h \right)} \rect_{[-\Tcp,0)} (i \Ts \! + \! \Tcp \! - \!\tau_h)}_{\color{Dark2-F}r^{\text{TD-CC}}_{\text{ISI},h,m}} \\
        + \underbrace{\sum_{n=1}^{N} X[n,m \! - \! 1] \e^{\j 2 \pi n \Delta f \left( i \Ts + T_0 - \tau_h \right)} g (i \Ts \! - \! T \! - \! \tau_h)}_{\color{Dark2-A}r^{\text{TD-CC}}_{\text{ISI},h,m\!-\!1}}.
    \label{eq:ch2:r_ISI}
    \end{gathered}
\end{equation}
Equation~(\ref{eq:ch2:r_ISI}) shows that additional \ac{ISI} stemming from the \mbox{$m$th} and \mbox{$(m\!+\!1)$th} transmit symbol is introduced, where the \ac{ISI} stems mainly from the \mbox{$(m\!+\!1)$th} transmit \ac{OFDM} symbol for a close target. Additionally, the \ac{TD}-\ac{CC} doubles the thermal noise power and Tab.~\ref{tab:ch2:power_level} summarizes the power levels.\\
The \ac{CC} can be also carried out in the frequency domain by
\begin{equation}
    Y^{\text{FD-CC}}[n,m] = Y[n,m] + C[n] \odot Y[n,m \!+ \!1],
    \label{eq:ch2:FD-CC}
\end{equation}
where $C[n] = \e^{-\j 2 \pi n \Ncp/N}$ compensates for the time delay introduced by the \ac{CP}. Note that \ac{FD}-\ac{CC} performs slightly worse than \ac{TD}-\ac{CC} because the signal which arrives during the \mbox{$(m\!+\!1)$th} cyclic prefix is lost. This means that a residual \ac{ICI} remains, as shown in Tab.~\ref{tab:ch2:power_level}.

\subsection{Coherent Compensation in a Multi Target Scenario}
\begin{figure}[!b]\flushleft
    \vspace{-0.7cm}
    \centering
    \input{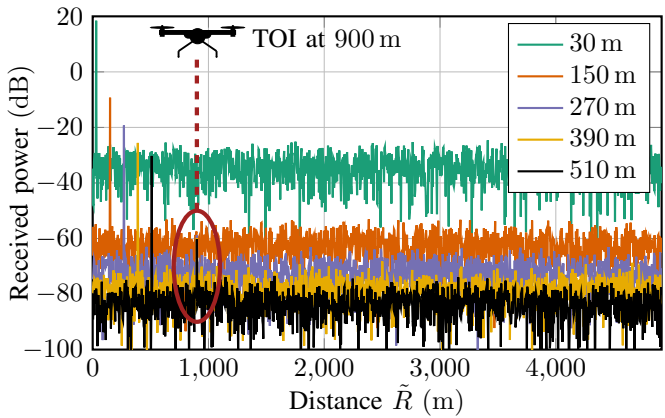}
    \vspace{-0.8cm}
    \caption{\ac{RDM} cut at $\tilde{v} = 0$ after the \ac{TD}-\ac{CC} for a scenario with two static targets. The range of the first target is given by the legend, and the distance of the second target is fixed at $R_2 = \SI{900}{\metre}$. The \ac{RCS} of the first and second target are $\sigma_{\mathrm{RCS},1} = \SI{10}{\metre\squared}$ and $\sigma_{\mathrm{RCS},2} = \SI{0.1}{\metre\squared}$ representing a truck and a drone, respectively. The \ac{ISI} introduced by the first target becomes stronger as it gets closer, preventing the detection of the second target.
    }
    \label{fig:ch3:noise:floor}
\end{figure}

\vspace{-0.05cm}
\label{sec:ch3:CC_in_MT}
To illustrate the drawback of the \ac{TD}-\ac{CC} in a multi target scenario, we consider a scenario with one close and one distant target, as depicted in Fig.~\ref{fig:ch1:scenario}. In conventional radar processing, only the distant target introduces \ac{ISI} (see Sec.~\ref{sec:ch2:target_exceeding_CP}). However, both targets introduce \ac{ISI} when the \ac{TD}-\ac{CC} is applied. As illustrated in Fig.~\ref{fig:ch2:Add_compensation}, the ($m \! + \! 1$)th receive symbol contains not only the $m$th transmit symbol from the reflection of the distant target but also, predominantly, the ($m\!+\!1$)th transmit symbol from reflections of the close target. Since this consecutive receive symbol is added to the actual receive symbol to restore the peak power of the distant target, \ac{ISI} from the $(m\!+\!1)$th transmit symbol is introduced into the $m$th symbol due to the close target (\ref{eq:ch2:r_ISI}). Given the reflected power decreases quartically with distance and the \ac{ISI} power scales linearly with the reflected power $a^2_h$ of the target, the \ac{ISI} from the close target is dominant because it is usually much stronger than that of the distant target. Figure~\ref{fig:ch3:noise:floor} shows a Doppler cut of the \ac{RDM} after \ac{CC} with various positions for the close target. We observe that this \ac{ISI} manifests as a floor, which increases as the first target comes closer, preventing the detection of the distant target. This renders scenarios with both a close and distant target highly challenging. Therefore, we will consider these scenarios in the following. Further, Fig.~\ref{fig:ch3:noise:floor} validates that the \ac{ISI} is mainly caused by the close target in this multi target scenario.

\subsection{Multi Target Coherent Compensation}
\ac{TD}-\ac{CC} enhances the peak power of distant targets~\cite{wang_coherent_2023} but the overall performance deteriorates in multi target scenarios due to an increased interference-and-noise floor. To address this limitation, we propose a \ac{MTCC}, a modified version of the \ac{FD}-\ac{CC}, which increases the peak power of the distant targets with only a minimal increase of the floor, ensuring a robust performance in multi target scenarios.

For the \ac{MTCC}, we first compute the \ac{RDM} without applying any \ac{CC}, as detailed in Sec.~\ref{sec:ch2:receiver}. Distant targets with small \acp{RCS}, which are of particular interest, may not be detected in this initial \ac{RDM} using conventional radar processing. However, we note that in this initial \ac{RDM} the close targets and the targets with a large \ac{RCS} are detectable. We recall that these targets will generate strong \ac{ISI} if the \ac{CC} is applied.

Since the \ac{ISI} after applying the \ac{CC} scales linearly with the peak power $a^2_h$ of the target (see Tab.~\ref{tab:ch2:power_level}), we propose to limit the peak power of the targets in the \ac{RDM} to a \mbox{threshold $\varepsilon \! \in \mathbb{R}^{+}$}
\vspace{-0.2cm}
\begin{equation}
    \mathrm{RDM}_{\mathrm{T}}[\tilde{R},\tilde{v}] = \begin{cases}
        \RDM [\tilde{R},\tilde{v}] , & | \RDM [\tilde{R},\tilde{v}] |^2 < \varepsilon \\
        \sqrt{\varepsilon}, & | \RDM [\tilde{R},\tilde{v}] |^2 \geqslant \varepsilon \\
    \end{cases}
    \label{eq:thresholding}
    \vspace{-0.1cm}
\end{equation}
to reduce the amount of \ac{ISI} introduced by the \ac{CC}. To avoid significant degradation of the image \ac{SINR}, the \ac{ISI} introduced by the \ac{CC} should be below the thermal noise floor. Since the \mbox{threshold $\varepsilon$} limits the \ac{ISI} to \mbox{$\varepsilon / \Gp $} per target (see Tab.~\ref{tab:ch2:power_level}) if the \ac{CC} is applied, $\varepsilon$ should be chosen such that $\varepsilon < \sigma^2_{\mathrm{Therm}} G_P$. However, it should be above the observed floor from the initial estimate, since it contains the signal components from the previous \ac{OFDM} symbol.

Next, the \ac{RDM} $\mathrm{RDM}_{\mathrm{T}}$ with attenuated peaks is transformed back into the frequency domain, and the corresponding received signal with attenuated peaks is derived
\vspace{-0.1cm}
\begin{equation}
    Y_{\mathrm{T}}[n,m] = \underset{N \downarrow}{\mathrm{FFT}} \left\{  \underset{M \rightarrow}{\mathrm{IFFT}} \left\{ \mathrm{RDM}_{\mathrm{T}} [\tilde{R},\tilde{v}] \right\} \right\} \odot X[n,m].
    \vspace{-0.15cm}
\end{equation}

The derived received signal $Y_{\mathrm{T}}[n,m\!+\!1]$ is fed to the \ac{FD}-\ac{CC} defined in (\ref{eq:ch2:FD-CC}), where it replaces the received signal $Y[n,m\!+\!1]$ and yields $Y^{\mathrm{MTCC}}$. 
The final \ac{RDM} $\mathrm{RDM}^{\mathrm{MTCC}}$ is computed based on $Y^{\mathrm{MTCC}}$ as described in Sec.~\ref{sec:ch2:receiver}. 

The final \ac{RDM} $\mathrm{RDM}^{\mathrm{MTCC}}$ has attenuated peaks because it is based on $\mathrm{RDM}_{\mathrm{T}}$~(\ref{eq:thresholding}). Therefore, $\mathrm{RDM}^{\mathrm{MTCC}}$ is combined with the initial \ac{RDM} $\mathrm{RDM}$ to restore the height of the peaks. Our proposed \ac{MTCC} adds only a slight increase in complexity, with the \ac{FFT} efficiently implemented at a low level and otherwise requiring only pointwise operations, as summarized in Algorithm~\ref{alg}.

\algrenewcommand{\Return}{\State\algorithmicreturn~}
\begin{algorithm}
    \caption{\Acf{MTCC}}\label{alg}
    \textbf{Input:} $X[n,m]$, $Y[n,m]$, $H_{\text{win}}[n,m]$, $\varepsilon$
    \hspace*{-1cm}
    \begin{algorithmic}[1]
    
        \State $
            \mathrm{RDM}[\tilde{R},\tilde{v}] = \underset{M \rightarrow}{\mathrm{FFT}} \left\{ \underset{N \downarrow}{\mathrm{IFFT}} \left\{ \frac{Y[n,m]}{X[n,m]} \odot H_{\mathrm{win}} [n,m] \right\} \right\}
        $ %
        \State {$
            \mathrm{RDM}_{\mathrm{T}}[\tilde{R},\tilde{v}] = \begin{cases}
            \RDM [\tilde{R},\tilde{v}] , & | \RDM [\tilde{R},\tilde{v}] |^2 < \varepsilon \\
            \sqrt{\varepsilon}, & | \RDM [\tilde{R},\tilde{v}] |^2 \geqslant \varepsilon \\
        \end{cases}
        $} %
        \State 
        $
            Y_{\mathrm{T}}[n,m] = \underset{N \downarrow}{\mathrm{FFT}} \left\{  \underset{M \rightarrow}{\mathrm{IFFT}} \left\{ \mathrm{RDM}_{\mathrm{T}} [\tilde{R},\tilde{v}] \right\} \right\} \odot X[n,m]
        $ %
        \State $
            Y^{\mathrm{MTCC}}[n,m] = Y[n,m] + C[n] \odot Y_{\mathrm{T}}[n,m\!+\!1]
        $ %
        \State $
            \mathrm{RDM}^{\text{MTCC}}[\tilde{R},\tilde{v}] = \underset{M \rightarrow}{\mathrm{FFT}} \left\{ \underset{N \downarrow}{\mathrm{IFFT}} \left\{ \frac{Y^{\mathrm{MTCC}}[n,m]}{X[n,m]} \right\} \right\}
            $
        \State $
            \mathrm{RDM}^{\text{MTCC}}[\tilde{R},\tilde{v}] \! \! = \! \! \begin{cases}
            \! \RDM^{\text{MTCC}} [\tilde{R},\tilde{v}] , &  \! \! \! \! \! | \RDM [\tilde{R},\tilde{v}] |^2  \! < \! \varepsilon \\
            \! \RDM [\tilde{R},\tilde{v}] , & \! \! \! \! \! | \RDM [\tilde{R},\tilde{v}] |^2  \! \geqslant \!  \varepsilon \\
        \end{cases}
        $
    \end{algorithmic}
    \textbf{Return:} $\mathrm{RDM}^{\mathrm{MTCC}}[\tilde{R},\tilde{v}]$
\end{algorithm}
\vspace{-0.5cm}

\subsection{Performance Metrics}
We use the image \ac{SINR} to assess the performance gain of our \ac{MTCC}. We generalize the image signal-to-noise ratio \cite{nuss_limitations_2018} to include also the quantization noise, \ac{ISI} and \ac{ICI} stemming from targets exceeding the \ac{ISI}-free range. The \ac{ISI} and \ac{ICI} manifest themselves as additive white Gaussian noise \cite{seoane_analysis_1997} and reduce the image \ac{SINR}, which is defined for a \ac{TOI} as
\begin{equation}
    \mathrm{Image \, SINR}_{\mathrm{TOI}} = \frac{P_{\mathrm{Peak},\mathrm{TOI}}}{P_{\mathrm{Therm}} \! + \! P_{\mathrm{Q}} \! + \! \sum_{h=1}^{H} \left( P_{\mathrm{ISI},h} \! + \! P_{\mathrm{ICI},h} \right) },
    \label{eq:ch3:image_snr}
\end{equation}
where the quantisation noise power is \cite{mandelli_survey_2023}
\begin{equation}
    P_{\mathrm{Q}} = \frac{P_{\mathrm{Sig}} \cdot \mathrm{PAPR}}{G_\mathrm{P} \cdot 10 \log_{10} \left( 6.02 Q\right)  },
    \label{eq:ch5:quantisation_noise}
\end{equation}
and $P_{\mathrm{Sig}}$ and $\mathrm{PAPR}$ denote the average signal power and peak-to-average power ratio of the received signal $r(t)$, respectively.

\section{Simulation Results}
\label{sec:simulation}

\begin{table}[!b]
    \vspace{-0.8cm}
    \caption{Considered system parametrizations \cite{mandelli_survey_2023}}
    \vspace{-0.2cm}
    \label{tab:ch5:system_parameter}
    \centering
    \begin{tabular}{cccc}
        \toprule
        Parameter & Symbol & Full frame & Partial frame \\
        \midrule
        Carrier frequency & $f_\mathrm{c}$ & \num{3.5} GHz & \num{3.5} GHz \\
        Bandwidth & $B$ & \num{200} MHz & \num{200} MHz  \\
        Sub-carrier spacing  & $\Delta f$ & \num{30} kHz & \num{30} kHz  \\
        No. sub-carrier & $N$ & \num{6552} & \num{6552} \\
        \acs{CP} length (sample) & $N_{\mathrm{cp}}$ & \num{458} & \num{458}  \\
        No. \ac{OFDM} symbols & $M$ & \num{280} & \num{10} \\
        Transmit power & $P_{\mathrm{Tx}}$ & \num{49} dBm & \num{49} dBm \\
        Antenna gain & $G_{\mathrm{Tx}}, G_{\mathrm{Rx}}$ & \num{25.8} dBi & \num{25.8} dBi \\
        Noise figure & $ F$  & \num{8} dB & \num{8} dB \\
        \bottomrule
    \end{tabular}
\end{table}

In this section, we compare conventional radar processing (no-\acs{CC})~\cite{braun_ofdm_2014}, the generalized state-of-the-art \acs{TD}-\acs{CC}~\cite{wang_coherent_2023} and our proposed \ac{MTCC} with a threshold of \mbox{$\varepsilon = 100P_{\mathrm{Therm}}$}. Throughout the simulations, the modulation alphabet is QPSK, and the assumed \ac{TOI} is a drone with a \ac{RCS} of $\SI{0.1}{\metre\squared}$. For simplicity, we assume that the distance is a multiple integer of the range resolution and that the receiver applies a rectangular window $H_{\mathrm{win}}[n,m] = 1$.

\subsection{Coherent Compensation in a Single Target Scenario}
\label{sec:rest:CC-ST}
\vspace{-0.1cm}

\begin{figure}
    \vspace{-0.2cm}
    \centering
    \input{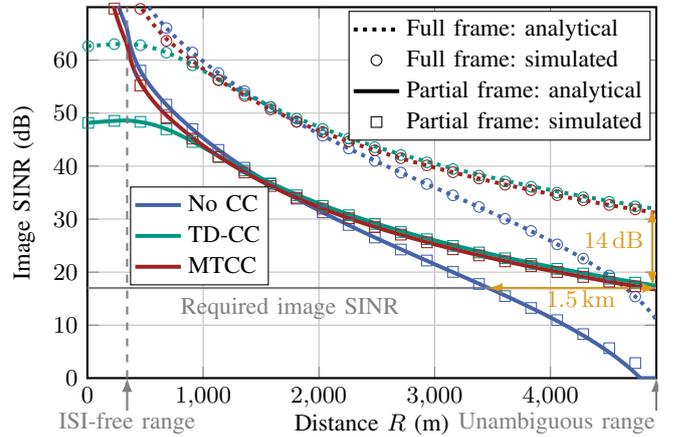}
    \vspace{-0.4cm}
    \caption{Single target scenario: Image \ac{SINR} as a function of its distance for the system parametrizations given in Tab.~\ref{tab:ch5:system_parameter} with and without \acs{CC}. For reliable detection, an image \ac{SINR} of at least $\SI{17}{dB}$ is required~\cite{mandelli_survey_2023}.
    }
    \label{fig:ch4:parameter_study}
    \vspace{-0.50cm}
\end{figure}
We first quantify the image \ac{SINR} improvement of a \ac{CC} in a scenario where only a single distant drone should be detected. Figure~\ref{fig:ch4:parameter_study} shows the analytically derived and simulated image \ac{SINR} as a function of the distance up to the unambiguous range \mbox{$R_{\text{max}} = N c_0/(2B)$}, which is $\SI{4914}{\metre}$ for our parameterizations. Tab.~\ref{tab:ch5:system_parameter} shows the two system parameterizations under investigation, a full frame and a partial frame using only 3.6\% of the \ac{OFDM} symbols for sensing~\cite{mandelli_survey_2023}. For reliable performance in terms of both false alarms and missed detection, an image \ac{SINR} of $\SI{17}{dB}$ is required \cite{mandelli_survey_2023}. We observe that a drone can generally be detected beyond the \ac{ISI}-free range. However, the image \ac{SINR} falls quickly below $\SI{17}{dB}$, which would result in a missed detection for long distances without a \acs{CC} mainly due to the reduced peak power (see Sec.~\ref{sec:ch2:target_exceeding_CP}). On the contrary, our \ac{MTCC} outperforms the classical \ac{OFDM} radar processing after $\SI{1750}{\metre}$, increases the drone detection distance up to the maximum range for both parametrizations, and improves the image \ac{SINR} up to $\SI{14}{dB}$ at the unambiguous range. This means that the partial frame suffices for drone detection up to the unambiguous range and only 3.6\% of the processing gains are required using the \ac{MTCC} compared to no \ac{CC}, resulting in 96.4\% less overhead in terms of radio resources required.

\begin{figure}[!t]
    \centering
    \input{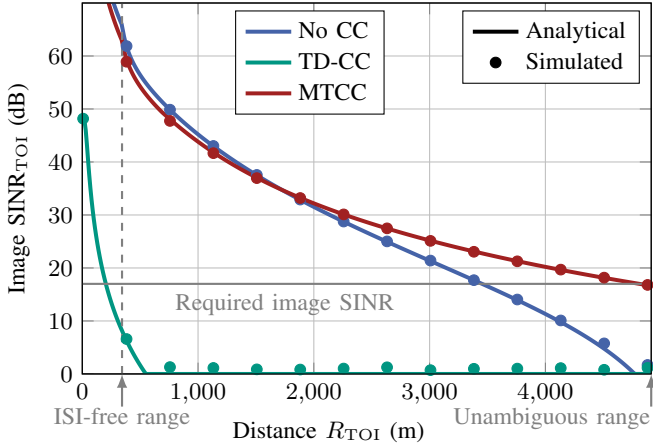}
    \caption{Multi target scenario: Image \ac{SINR} of the \ac{TOI} as a function of its distance for the partial frame parametrization given in Tab.~\ref{tab:ch5:system_parameter} with and without \acs{CC}. The distance of the second, interfering target is fixed at $R_{\text{Int}} = \SI{105}{\metre}$.}
    \label{fig:ch4:Range_vs_Image_SNR}
\end{figure}

\vspace{-0.1cm}

\subsection{Coherent Compensation in a Multi Target Scenario}
\vspace{-0.1cm}
Next, we add a second, interfering target close to the base station at \mbox{$R_{\text{Int}} = \SI{105}{m}$} with \mbox{$\sigma_{\mathrm{RCS},\text{Int}} = \SI{10}{\metre\squared}$}, representing, e.g., a truck, to evaluate the performance of our algorithm in a multi target scenario. Figure~\ref{fig:ch4:Range_vs_Image_SNR} depicts the image \ac{SINR} as a function of the range of the \ac{TOI}, the drone. As in the single target scenario, distant targets cannot be detected using conventional \ac{OFDM} radar processing. The image \ac{SINR} after \ac{TD}-\ac{CC} degrades rapidly due to the \ac{ISI} introduced by the second, interfering target (see Sec.~\ref{sec:ch3:CC_in_MT}). On the contrary, our \ac{MTCC} achieves the same \ac{SINR} as in the single target scenario. This demonstrates that our proposed \ac{MTCC} increases the peak power of the target without increasing the \ac{ISI} power significantly enabling reliable long-range sensing in multi target scenarios.

\subsection{Scenario Analysis} %
\begin{figure}[!t]
    \centering
    \input{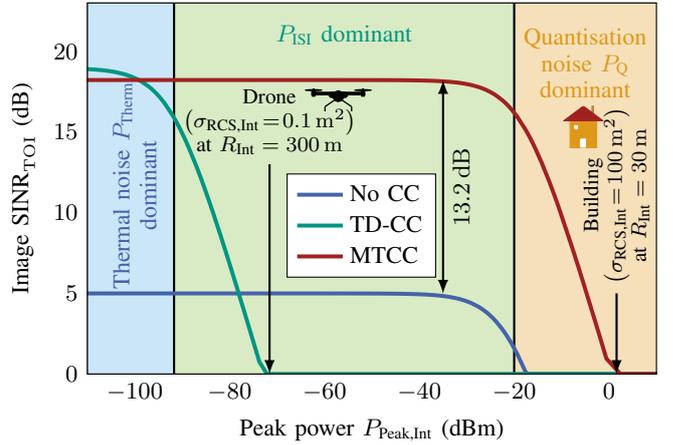}
    \vspace{-0.2cm}
    \caption{Image \ac{SINR} of the \ac{TOI} (drone) as a function of the peak power of the interfering target. Specifically, a peak power $P_{\text{Peak,Int}}$ of $\SI{-71.6}{dBm}$ and $\SI{1.6}{dBm}$ corresponds to a drone \mbox{$\left( \sigma_{\mathrm{RCS},\text{Int}} = \SI{0.1}{\metre\squared} \right)$} at a distance of \mbox{$R_{\text{Int}} = \SI{300}{\metre}$} and a building \mbox{$\left( \sigma_{\mathrm{RCS},\text{Int}} = \SI{100}{\metre\squared} \right)$} at a distance of \mbox{$R_{\text{Int}} = \SI{30}{\metre}$}, respectively.}
    \label{fig:ch3:operating_points}
    \vspace{-0.2cm}
\end{figure}

Finally, we consider a scenario in which the \ac{TOI} has a fixed distance \mbox{$R_{\text{TOI}} = \SI{4500}{m}$}. We sweep the distance and \ac{RCS} of the interfering target to demonstrate the wide performance improvement of our proposed \ac{MTCC}. Figure~\ref{fig:ch3:operating_points} shows the image \ac{SINR} of the \ac{TOI} as a function of the peak power of the interfering target $P_{\text{Peak,Int}}$.

Recall that \ac{TD}-\ac{CC} introduces \ac{ISI}, which is $G_{\mathrm{P}}$ times weaker than the peak power. As indicated by the blue area in Fig.~\ref{fig:ch3:operating_points}, if the peak power of the interfering target is small, the \ac{ISI} introduced by \ac{CC} is less than the thermal noise power, causing only minor performance degradation without the need for adaptation of the \ac{CC}. Conversely, in the presence of an exceedingly strong interfering target, the quantization noise constrains the detection as indicated by the orange area in Fig.~\ref{fig:ch3:operating_points}. Therefore, detecting the secondary target is not feasible even with our \ac{MTCC}. However, note that our \ac{MTCC} is more robust to quantization noise compared to no \ac{CC} because the peak power of the \ac{TOI} is larger (see Tab.~\ref{tab:ch2:power_level}). If the peak power of the interferer lies between these two extremes, our method improves the image \ac{SINR} by \SI{13.2}{dB} compared to no \ac{CC}, (due to reduced peak power) and enables reliable detection, where \ac{TD}-\ac{CC} fails (due to strong additional \ac{ISI}). Therefore, Fig.~\ref{fig:ch3:operating_points} demonstrates the superior and robust performance of our proposed \ac{MTCC} over variations up to $\SI{70}{dB}$ in the peak power of the interfering target, covering most realistic scenarios, where drones and buildings are interfering targets.

\section{Conclusion}
\Ac{CP}-\ac{OFDM}-\ac{ISAC} systems suffer from a significant loss in image \ac{SINR} when detecting distant objects because their corresponding path delays surpass the \ac{CP} duration. We showed that existing algorithms are not suitable for multi target scenarios due to strong additional \ac{ISI}. 
As a solution, we proposed \ac{MTCC}, a novel and robust algorithm maintaining the image \ac{SINR} in multi target scenarios. We derived analytical expressions for the image \ac{SINR} and verified them through numerical simulations. Our proposed \ac{MTCC} increases the image \ac{SINR} by up to $\SI{14}{dB}$ at long ranges. This allows less overhead of required radio resources to generate enough processing gain for reliable target detection, with up to 96.4\% reductions in our numerical study. Our proposed \ac{MTCC} paves the way for resource-efficient \ac{6G} \ac{ISAC} systems that can master challenging tasks like long-range drone detection.

\end{document}